STOLZENBERG'S "THE HOLY OFFICE IN THE REPUBLIC OF LETTERS" REVISITED: ON AN ASTRONOMICAL DIAGRAM AND WHETHER THE PAPACY TACITLY PERMITTED THE CIRCULATION OF AN EXPLICITLY COPERNICAN BOOK IN 1660

Christopher M. Graney
Vatican Observatory
00120 Stato Città del Vaticano
c.graney@vaticanobservatory.org

**Abstract**

**Did the papacy tacitly permit the circulation of an explicitly Copernican book in 1660? One scholar has recently argued that it did. A close analysis of a unique illustration from that book, Andreas Cellarius's atlas *Harmonia Macrocosmica*, illuminates this argument. This is because the illustration, a diagram showing the relative sizes of the sun, moon, planets, and stars, was among the material reviewed (at the request of the book's publisher) by the Holy Office prior to the book's publication and was pro-Copernican.**

**Introduction**

The Amsterdam bookseller Johannes Janssonius and the Roman Inquisition quietly collaborated in 1660 to enable an explicitly Copernican book to circulate in Italy. So argues Daniel Stolzenberg in "The Holy Office in the Republic of Letters: Roman Censorship, Dutch Atlases, and the European Information Order, circa 1660", published in *Isis* in 2019. The book in question was Andreas Cellarius's *Harmonia Macrocosmica, seu Atlas Universalis et Novus*, or *Celestial Atlas*. It was "blatantly pro-Copernican", says Stolzenberg, indeed "a partisan defense of Copernicanism". This paper provides additional details about the contents of the atlas, centered on one of its illustrations. These details support the idea that "the papacy tacitly permitted the circulation of an explicitly Copernican book at a surprisingly early date".[1]

---

[1] Daniel Stolzenberg, "The Holy Office in the Republic of Letters: Roman Censorship, Dutch Atlases, and the European Information Order, circa 1660," *Isis*, 110 (2019), 1-23, quotations here on 8, 10, 1.

Specifically, the *Harmonia Macrocosmica* employed a little scientific sleight-of-hand regarding the various bodies in the universe (Earth, sun, moon, planets, and stars) to mask what in 1660 was considered a great scientific weakness of Copernicus's sun-centered universe. Cellarius made his book ostensibly *not* pro-Copernican by prominently labelling Copernicus's world system as a "hypothesis" in the book's illustrations. The sleight-of-hand, by contrast, was subtle. It would have been most likely to be noticed by knowledgeable people familiar with the technical details of the debates over the various "hypotheses" that had been put forth regarding how the universe was structured. Two such people were the men eventually tasked by the Holy Office with the review of Cellarius's book. These men gave their approval to the book, ignoring not only Cellarius's sleight-of-hand but also a long-standing question of faith and science that his illustration raised.

**The "Collaboration", in Brief**

A synopsis of "The Holy Office in the Republic of Letters" is now needed.

In 1660 Elizeus Weyerstraet, Janssonius's agent, arrived at Rome on a sales trip. Along with much else, he was carrying material for the yet-to-be-published *Harmonia Macrocosmica*. He presented this material to the Holy Office on his own initiative. "Weyerstraet provided the Inquisition with copies of its [i.e. *Harmonia Macrocosmica's*] engraved illustrations [Figure 1], which he asked them to review so that Janssonius could ensure that they would not raise hackles with an important market sector. The Holy Office obliged... instructing its agents to proceed quickly and to treat the Dutch bookseller kindly (*humaniter*)."[2] Janssonius was eager that the atlas sell in Catholic countries. Rome was eager to kindly oblige because at that time Pope Alexander VII was working to use "soft power" after the Peace of Westphalia, conducting a "missionary charm offensive" to reach out to and win over prominent Protestants. Furthermore, "at all levels the Curia was staffed by members of the Republic of Letters" who had broad interests and connections that reached beyond the Catholic world.[3]

The two men tasked with the review of the Cellarius illustrations both had close ties to Alexander VII. Michelangelo Ricci had been a student of Galileo's friend Benedetto Castelli; had been a collaborator with Galileo's successor at the Tuscan court, Evangelista Torricelli; and was author of a treatise on geometry.[4] Athanasius Kircher, S.J. of the Roman College was author of many books pertaining to mathematics and natural philosophy. In Cellarius's illustrations, the Copernican system of the universe, and other systems, were labelled as "hypotheses". Both reviewers found acceptable this presentation of hypotheses with no claim of truth; both assessed the illustrations as containing nothing "contrary to faith".[5] "After receiving Ricci's and Kircher's judgments, the Holy Office decreed that, although it would still be necessary to see the texts [that would accompany the illustrations in the atlas], so long as they did not contain anything 'absurd' (in 1616 the Holy Office had declared heliocentrism to be 'philosophically absurd') the atlas would by no means be prohibited."[6]

---

[2] *Ibid.*, 4.
[3] *Ibid.*, 16-17.

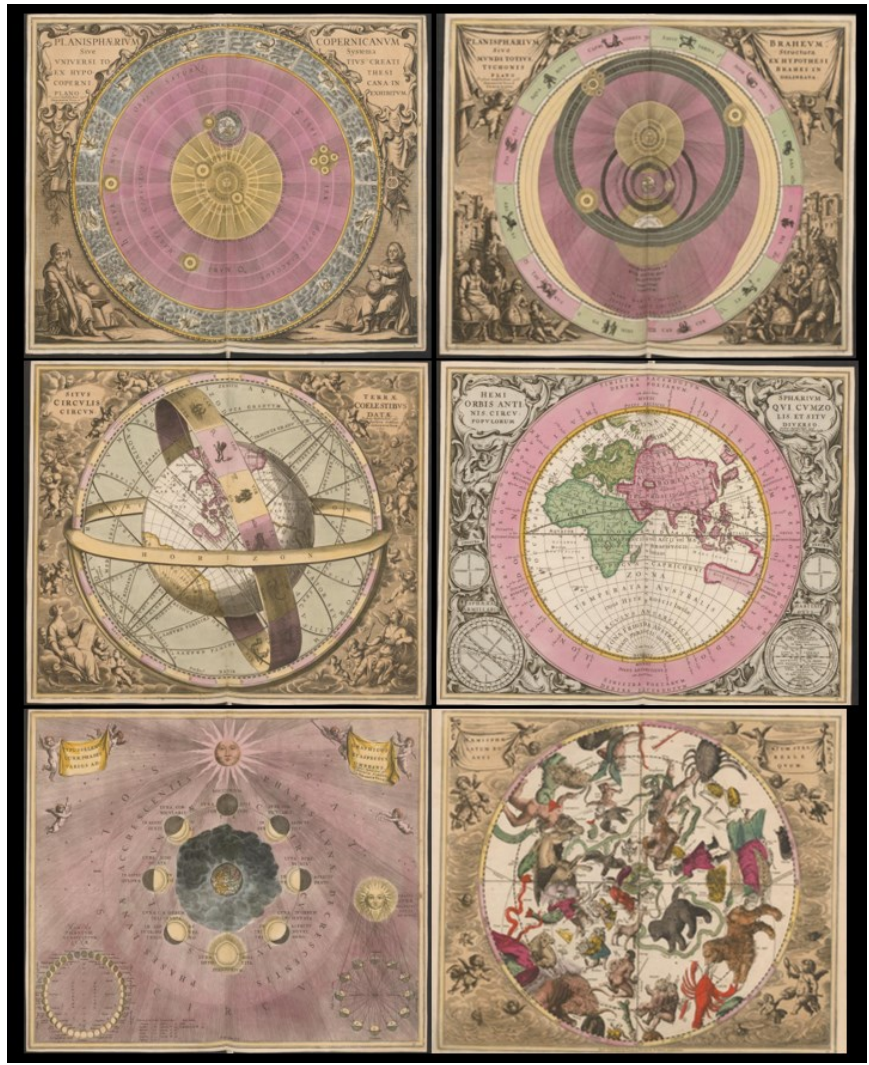

**Figure 1: A selection of illustrations from Andreas Cellarius's *Harmonia Macrocosmica*, from a 1708 reprint. Top two are the hypotheses of Copernicus (left) and Tycho Brahe (right). All Cellarius images in this paper are credit ETH-Bibliothek Zürich (public domain mark).**

---

[4] Ricci's treatise was less than 20 pages long. He became a cardinal in 1681.
[5] *Ibid.*, 5-6.

The texts, however, turned out to be pro-Copernican. The atlas's frontispiece, which like the texts was not reviewed, was likewise pro-Copernican (Figure 2). Nevertheless, the prologue of the atlas claimed that Cellarius was simply showing the various hypotheses and taking sides. It seems this declaration was enough. Janssonius shipped a crate of copies of *Harmonia Macrocosmica* to Italy in 1661, with one copy specifically designated for the Vatican Library. No problems arose. What is more, in 1661 Janssonius struck a deal with Kircher in which Kircher would be well-paid for the rights to print and sell all his past and future works. And so "The Copernican *Celestial Atlas* circulated in Italy because of, not despite, the Holy Office."[7]

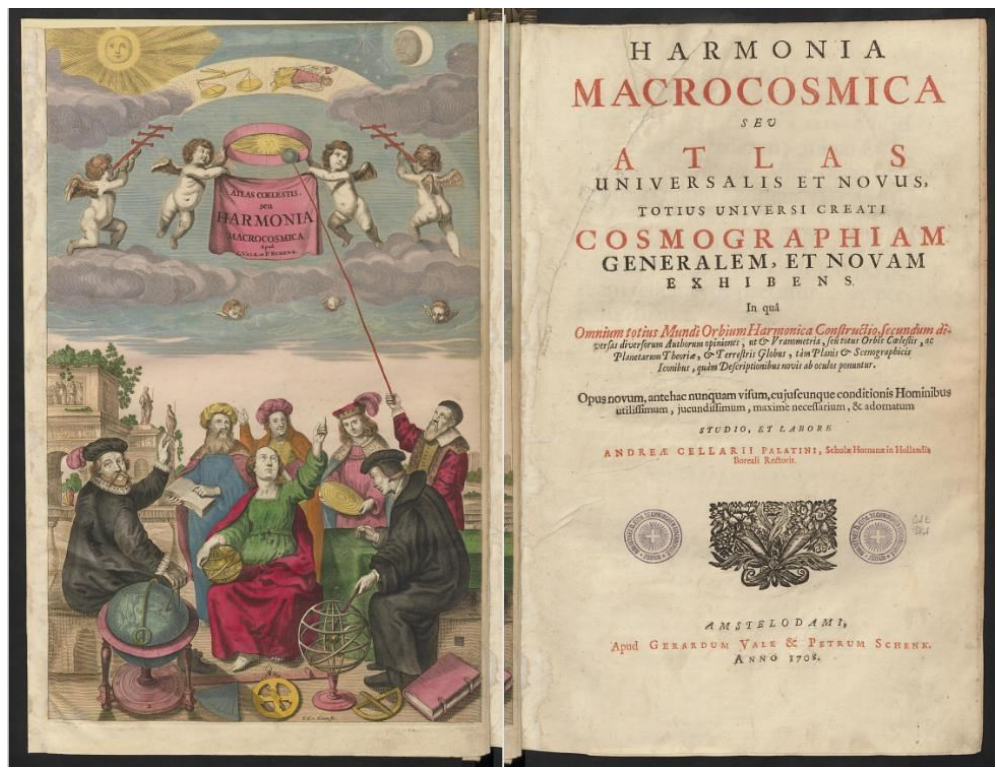

**Figure 2:** *Harmonia Macrocosmica* frontispiece and title page. Note the Earth circling the sun at top and the Copernican diagram held by the third figure from the right.

---

[6] *Ibid.*, 17.
[7] *Ibid.*, 8-10, 16, 18, 23 (quotation).

**Sizes of Celestial Bodies**

That a Copernican work circulated in 1660 *because of* the Holy Office is a bold statement. Indeed, given that the frontispiece and text of Cellarius's atlas were not reviewed by Kircher and Ricci, one might object that it is too bold. One might suppose instead that the reason no problems arose when the atlas arrived in Italy is that no one actually read the published book; the copy sent to the Vatican library sat on a shelf there; Roman authorities never bothered to make sure that the atlas's texts contained nothing absurd.[8]

That possible objection is significantly weakened by a careful study of one of the illustrations that Weyerstraet provided to the Holy Office—an illustration labelled simply, "Corporum Coelestium Magnitudines" or "Sizes of Celestial Bodies", with little other text (Figure 3). "Corporum" provides a comparison of the sizes of various bodies, to each other and to a scale of German miles and Earth diameters. The bodies pictured include the sun, moon, Earth, the five planets known in 1660 (those visible to the unaided human eye: Mercury, Venus, Mars, Jupiter and Saturn), and a representative star from each of the six classes or "magnitudes" of the stars.

These classes indicated the prominence of stars. They had been used by astronomers since antiquity. The term "magnitude" spoke to the apparent "largeness" of stars. The most prominent stars in the sky were called stars of the "first magnitude", what one might think of as "first class stars". Stars barely visible to the keen unaided eye under a dark, clear sky (ideal weather and no moon or non-celestial light sources present) were of the sixth magnitude. Stars in between first and sixth were classified with some disagreement as to where one class ended and the other began.

Progressing from sixth up to first magnitude, stars became both larger in apparent size and less numerous. There were a handful of well-known first magnitude stars. Various astronomers from Ptolemy in the second century A.D. down to Tycho Brahe in the late sixteenth century all estimated these to measure approximately two-thirtieths the apparent diameter of the moon (or the sun, whose apparent size is roughly equal to that of the moon, as seen in a solar eclipse). There was a plethora of fifth and sixth magnitude stars with considerably smaller apparent sizes.[9]

---

[8] Stolzenberg considers this, stating "it must be acknowledged that there is no direct evidence that [Roman] officials… ever became aware that the text of the atlas violated their decrees", but argues that it is "highly probable" that officials were aware (*Ibid.*, 8-11, quotations on 11).

Figure 3: Top—"Corporum Coelestium Magnitudines". Bottom—detail showing globes of the sun, a first magnitude star, Jupiter, etc., with the smallest globes being the moon and Mercury. Arrows indicate five and four terrestrial diameters.

The differing apparent sizes of stars could be attributed to differing distances (sixth magnitude stars being farther away than first magnitude stars) or differing physical sizes (sixth magnitude stars being of lesser bulk than first magnitude stars). "Corporum" presumes the latter. It shows a first magnitude star as measuring 4 & 3/4 times the diameter of Earth, a sixth magnitude star measuring 2 & 5/8 Earth diameters.

**Not Copernican, but pro-Copernican nonetheless**

These sizes are arguably a bit of pro-Copernican scientific sleight-of-hand. They are problematic for a book produced in 1660. They are not based on a Copernican universe, and they are not based on telescopic measurements; Cellarius cites a pre-telescopic, non-Copernican source for the numbers used to produce the "Corporum" diagram: Christopher Clavius's (S.J.) commentary on a standard astronomy text, *The Sphere*. The numbers used match those found in a 1601 edition of Clavius.[10] *Harmonia Macrocosmica's* illustration of the physical sizes of celestial bodies shows those sizes as calculated six decades earlier by a pre-telescopic astronomer for an Earth-centered universe.

Were all celestial bodies equidistant from Earth, arranged on a simple Earth-centered celestial sphere, then calculating their sizes would be simple, because apparent sizes would reflect physical sizes; the sun and moon appear larger than the stars to the Earth-bound human eye, and so they would be larger in terms of physical bulk as well. But basic observations and geometry showed the sun to be more remote than the moon. For example, the moon passes in front of the sun during a solar eclipse. The two bodies are not equidistant from Earth, so were they equal in physical size the sun would have the smaller apparent size. Since sun and moon are equal in apparent size, yet the sun is more distant than the moon, then the sun must be the larger in in physical size—larger in bulk.

---

[9] Albert Van Helden, *Measuring the Universe: Cosmic Dimensions from Aristarchus to Halley* (Chicago, 1985), 24-53.
[10] Christopher Clavius, *In Sphaeram Ioannis De Sacro Bosco Commentarius* (Venice,1601), 186; there were many editions of this work by Clavius—see James M. Lattis, *Between Copernicus and Galileo: Christoph Clavius and the Collapse of Ptolemaic Cosmology* (Chicago, 1994); Cellarius, *Harmonia Macrocosmica* (Amsterdam, 1708), pars prior 69.

Those same observations and geometry indicated the stars to be very distant. As Ptolemy stated in his *Almagest*, the Earth is but a point in comparison to the distance to the stars:

> Now, that the earth has sensibly the ratio of a point to its distance from the sphere of the so-called fixed stars gets great support from the fact that in all parts of the earth the sizes and angular distances of the stars at the same times appear everywhere equal and alike, for the observations of the same stars in the different latitudes are not found to differ in the least.[11]

For example, an observer at the equator who sees stars overhead, such as those in Orion's belt, will be closer to those stars, by a significant fraction of Earth's size, than an observer at mid-latitude (Figure 4). That closer perspective yields no difference in the appearance of Orion's belt; therefore that significant fraction of Earth's size must be as nothing compared to the distance of the stars. The vast stellar distances implied vast stellar bulks in order for stars to appear even as small as they do. Thus Cellarius's diagram shows even sixth magnitude stars, those barely visible to keen eyes under good conditions, to be considerably larger than Earth. It shows first magnitude stars to be comparable to the sun (Figure 5).

Cellarius's diagram also shows the stars to be comparable in physical size to the planets (Figure 5). To the unaided eye, a modestly bright star compares to Saturn in apparent size. In an Earth-centered universe, the stars can lie just beyond Saturn.[12] Since that modestly bright star and Saturn are similar in both apparent size and distance, they must be similar in physical size as well. Thus Cellarius shows first magnitude stars, Jupiter, Saturn, and second and third magnitude stars all to have diameters between five and four times that of Earth—a bit smaller in size than the sun.[13]

"Corporum" thus shows a harmonious picture of celestial sizes, with all the celestial bodies neatly arranged, one in front of another, from the largest, the sun, down to the smallest, the moon and Mercury. It is a bit of pro-Copernican sleight-of-hand because it does not identify these sizes as belonging to any particular "hypothesis", and because in the Copernican

---

[11] Ptolemy, "The Almagest I, 6" in *Great Books of The Western World (16): Ptolemy, Copernicus, Kepler* (Chicago, 1952), 10.
[12] Van Helden, *Measuring the Universe*, 27, 50.
[13] In diagram (see Figure 3), "Solis, omnium coelestium corporum maximi orbicularis circuitus, et magnitudo". For numbers, see Cellarius, pars prior 69.

C. M. Graney – page 8 of 15

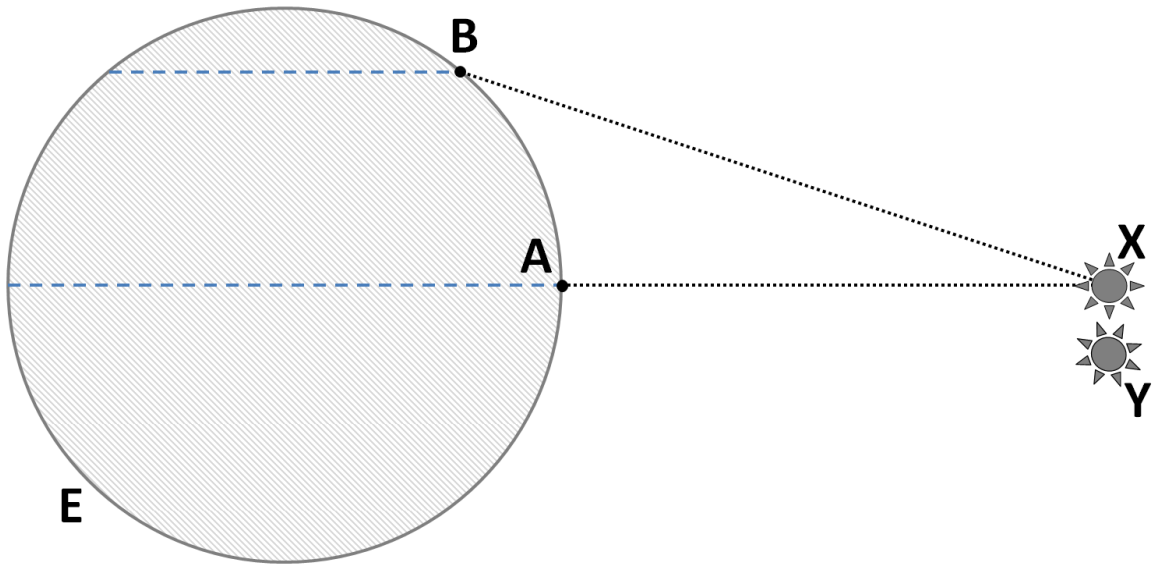

**Figure 4:** Circle E is the globe of the Earth. An observer A on the equator sees star X directly overhead. X is more distant from an observer B at a latitude away from the equator, and thus B must see X as smaller and dimmer than does A. B must also see the separation of stars X and Y differently than does A. Ptolemy noted that observers at different latitudes do not see such differences, so the size of the Earth must be negligible compared to the distances to the stars; if E is reduced to a point, A and B will both see X and Y the same.

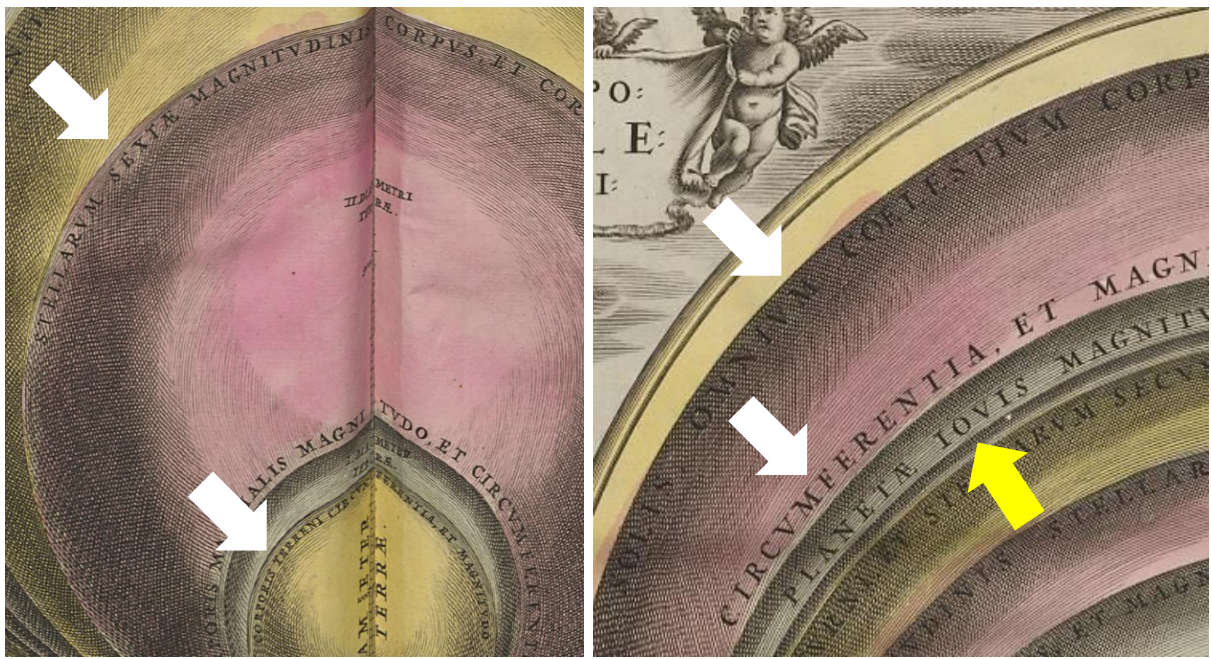

**Figure 5:** Left—a sixth magnitude star (upper arrow) and the Earth. Right—the sun (upper arrow), a first magnitude star, and Jupiter (yellow arrow).

hypothesis, the celestial size picture was not harmonious at all. Indeed, no such size diagram was possible for a Copernican universe.

In a Copernican universe, the fixed stars had to be located much farther away than Saturn. Ptolemy had stated that since the stars' appearances were independent of the place on Earth's globe from which they were viewed, the Earth must be like a point compared to their distances. In a Copernican universe, with the Earth in motion about the sun, the stars' appearances were independent of the place on Earth's *orbit* from which they were viewed. Moving from place to place on Earth's globe does not change the appearance of Orion's belt, and neither does moving from place to place on Earth's orbit (i.e. observing at different times of the year). Therefore, the Earth's *orbit* must be like a point compared to stellar distances.

Thus as Andreas Tacquet, S.J. (1612-1660) pointed out, whatever proportion existed in a geocentric universe between the size of Earth's *globe* and the sizes of the stars, that same proportion existed in a Copernican universe between the size of Earth's *orbit* and the sizes of the stars.[14] Consequently, to create a version of Cellarius's diagram for a Copernican universe, the scale of "terrestrial diameters" would have to become a scale of "orbital diameters"; first, second, and third magnitude stars would all have diameters between five and four times that of Earth's orbit rather than of Earth itself; the sun and planets in the diagram would be reduced to dots that would be tiny, if visible at all. Stars were giant in a Copernican universe, to the point that no representation of the relative sizes of the sun, stars, and planets in a Copernican universe would be possible in a single diagram. There could be no "Corporum Coelestium Magnitudines" for the Copernican hypothesis.

In the decades before the publication of *Harmonia Macrocosmica*, both Copernicans, like Johannes Kepler and Philips Lansbergen, and anti-Copernicans, like Tycho Brahe, Simon Marius, Christoph Scheiner, S.J., Giovanni Battista Riccioli, S.J., as well as Tacquet, recognized that stars in a Copernican universe would have to be giant. Telescopic observations did not alter this substantially, as both Marius and Riccioli took pains to point out. Kepler and Lansbergen welcomed the giant stars as manifestations of God's power.[15] But for the most part giant stars

---

[14] For more on Tacquet's reasoning, see Christopher M. Graney, "Galileo Between Jesuits: The Fault is in the Stars," *Catholic Historical Review* 107 (2021), 191-225.

[15] Anti-Copernicans generally accepted the hypothesis of Tycho Brahe, in which the sun, moon, and stars circled the Earth while the planets circled the sun—a hypothesis that was fully compatible with Galileo's telescopic discoveries that showed, for example, that Venus circled the sun. On Kepler and giant stars, see Christopher M. Graney, "As Big as a Universe: Johannes Kepler on the Immensities of Stars and of Divine Power," *Catholic*

were viewed as a problem for—an absurdity of—the Copernican hypothesis. Robert Hooke in 1674 would call the giant stars, "a grand objection alledged by divers of the great *Anti-copernicans* with great vehemency and insulting; amonst which we may reckon *Ricciolus* and *Tacquet*… hoping to make it [the Copernican system] seem so improbable, as to be rejected by all parties."[16] Even Galileo in his *Dialogue* had Sagredo (his "neutral observer") declare that for a star to be "so immense in bulk as to exceed the earth's orbit" is "a thing which is… entirely unbelievable."[17] Melchior Inchofer, S.J., who served on the special commission appointed by Pope Urban VIII to investigate the *Dialogue*, pointed to this grand anti-Copernican objection. Tacquet noted how Galileo tried and failed to argue against it.[18]

Cellarius, in opting to represent the relative sizes of celestial bodies using numbers from Clavius for a geocentric universe, rather than numbers from Kepler or Riccioli for a Copernican universe, obscured what many viewed as the Copernican theory's greatest weakness.[19] He also opted to not even associate the "Corporum" diagram with the geocentric hypothesis. In this way he steered his readers away from considering whether the sizes of celestial bodies might be different under different hypotheses. The diagram supports Stolzenberg's characterization of *Harmonia Macrocosmica* as being a partisan defense of Copernicanism.

**Not Easily Missed**

Ricci and Kircher did not review the pro-Copernican text and frontispiece of *Harmonia Macrocosmica* that led Stolzenberg to his conclusions. They did review the "Corporum" diagram. A knowledgeable reviewer familiar with the world system debate and on the lookout

---

*Historical Review* 105 (2019), 75-90; on Scheiner, see Graney, *Mathematical Disquisitions: The Booklet of Theses Immortalized by Galileo (*Notre Dame, IN, 2017); for the others, see Graney, *Setting Aside All Authority: Giovanni Battista Riccioli and the Science against Copernicus in the Age of Galileo* (Notre Dame, IN, 2015).

[16] Robert Hooke, *An Attempt to Prove the Motion of the Earth from Observations* (London, 1674), p. 26.

[17] Galileo Galilei, *Dialogue Concerning the Two Chief World Systems: Ptolemaic and Copernican*, trans. Stillman Drake [Modern Library Science] (New York, 2001), 432.

[18] André Tacquet, *Opera Mathematica* (Antverpiæ, 1668), 209: "Galilaeus in suo Mundi Systemate immanem istam Fixarum magnitudinem nequidquam conatur eludere [Galileo in his System of the World attempts in vain to elude this monstrous magnitude of the Fixeds by a long discourse]". Richard J. Blackwell, *Behind the Scenes at Galileo's Trial: Including the First English Translation of Melchior Inchofer's Tractatus Syllepticus* (Notre Dame, IN, 2006), 182: "the defenders of the Copernican system imagine that, since the stars are seen at an almost infinite distance, they have a size which is explicable by hardly any proportion". See also Graney, "Galileo Between Jesuits", 201-202, 216-17.

[19] This weakness would eventually be eliminated when the apparent diameters of stars, even those seen telescopically, were discovered to be spuriously large, mere artefacts of optics and the wave nature of light.

for absurdities would have seen the sleight-of-hand in "Corporum". Kircher certainly would have.

Kircher was familiar with Riccioli's work. He corresponded with Riccioli.[20] Moreover, in his writings Kircher cited Riccioli's 1651 *Almagestum Novum*, in which Riccioli laid out the star size problem complete with telescopic measurements.[21] One might try to suppose that Kircher simply missed that material in the *Almagestum Novum's* 1400-page vastness, but Kircher also cited Marius's 1614 *Mundus Iovialis*, in which Marius likewise presented the star size problem against the Copernican hypothesis using telescopic observations; *Mundus* is short.[22] So is the 1614 *Disquisitiones Mathematicae* of Scheiner, where Scheiner laid out the star size question in short form: the Earth's orbit is as a point in a Copernican universe; stars, having small but measurable diameters, are bigger than points; therefore, Copernican stars are bigger than Earth's orbit. Kircher at times borrowed heavily from *Disquisitiones*.[23] He criticized Scheiner's proposal in *Disquisitiones* (made well ahead of Newton) that Earth's orbit about the sun could be explained as a kind of perpetual fall.[24] It is hard to imagine how Kircher could not have seen Cellarius's sleight-of-hand.[25]

Readers of the published *Harmonia Macrocosmica* certainly could have seen it. They would not have had to delve deep into the text that lay beyond Cellarius's lengthy "taking no sides" prologue to see that the atlas's frontispiece was not its only pro-Copernican material. They would simply have had to look at the pictures and turn the pages until they happened upon "Corporum".

---

[20] Michael John Gorman, "The Angel and the Compass: Athanasius Kircher's Magnetic Geography," in *Athanasius Kircher: The Last Man Who Knew Everything*, ed. Paula Findlen (New York, 2004), 239-259, here 250-54.

[21] Athanasius Kircher, *Itinerarium Exstaticum* (Romæ, 1656), 9, 27, 463.

[22] Kircher, *Itinerarium*, 22, 464. An English translation of *Mundus* is available—see A. O. Prickard and Albert Van Helden, "The World of Jupiter, English Translation of Mundus Iovialis," in *Simon Marius and His Research*, eds. Hans Gaab and Pierre Leich (Cham, Switzerland, 2018), 1-54, here 8.

[23] See Athanasius Kircher, *Ars Magna Lucis et Vmbrae* (Romæ, 1645), 578, 756, 759 vs. Christoph Scheiner, *Disquisitiones Mathematicae* (Ingolstadii, 1614), 76, 76, 81. For an English translation of *Disquisitiones*, see Graney, *Mathematical Disquisitions*; the short form of the star size question is on page 30.

[24] Christopher M. Graney, "How to Make the Earth Orbit the Sun in 1614," *Journal for the History of Astronomy* 50 (2019), 16-30, here 27-28.

[25] It would be even harder to imagine if Kircher himself discussed giant Copernican stars. Ingrid Rowland mentions Kircher describing in his *Itinerarium Exstaticum* many stars "far larger than the Sun itself", but does not provide a specific reference, and I have been unable to find, in this work or in others of his, an unambiguous discussion of giant stars comparable to the discussions of Brahe, Scheiner, Kepler, etc. However, my searching has been by no means exhaustive. See Ingrid D. Rowland, "Athanasius Kircher, Giordano Bruno, and the Panspermia of the Infinite Universe," in Findlen, *Athanasius Kircher*, 191-205, here 196.

Here we should keep in mind that the "Corporum" diagram might catch the eyes of readers for reasons related not to the Copernican hypothesis, but to a far older instance of conflict between astronomy and scripture. Genesis 1:14-16 describes God as creating "the two great lights"—sun and moon—and the stars. But as we have seen, Ptolemy's astronomy showed quite persuasively that stars are in fact much larger than the moon, making the moon not so "great".

St. Augustine had addressed this. It may be true, he wrote in his *On the Literal Interpretation of Genesis*, that the stars are large, and merely "seem small because they have been set further away." But, he said, "at least grant this to our eyes... it is obvious that the Sun and Moon shine more brightly than the rest upon the Earth."[26] St. Thomas Aquinas offered the same opinion. "The two lights are called great, not so much with regard to their dimensions as to their influence and power," he wrote in *Summa Theologica*. "For though the stars be of greater bulk than the moon... as far as the senses are concerned, its apparent size is greater."[27] Genesis was not providing a scientific description of the relative bulks of the moon and stars.

This Genesis star size question had not been forgotten. John Calvin had addressed it his commentaries on scripture, making the same point as Augustine and Aquinas, but more vehemently. He expressed both enthusiasm for astronomy[28] and disdain for "janglers" who criticize Genesis saying that the moon is a great light. Genesis "does not call us up into heaven," he argued, but "only proposes things which lie open before our eyes."[29] "The Holy Spirit had no intention to teach astronomy; and, in proposing instruction meant to be common to the simplest and most uneducated persons, he made use... of popular language," he said.[30] Blaise Pascal had referenced Genesis and the question of the greatness of the moon in writing to François Annat, S.J. in 1657.[31]

---

[26] St. Augustine (Bishop of Hippo), *The Literal Meaning of Genesis*, in *The Works of Saint Augustine, a Translation for the 21st Century, Part I, Volume 13: On Genesis: A Refutation of the Manichees; Unfinished Literal Commentary on Genesis; The Literal Meaning of Genesis—introductions, translation and notes by Edmund Hill, O.P., editor John E. Rotele, O.S.A.* (Hyde Park, NY, 2002), 212 (Book II, 16.33).
[27] Question LXX ("Of the Work of Adornment, as regards the Fourth Day—In Three Articles") in *"The Summa Theologica" of St. Thomas Aquinas, Part I. QQ. I-LXXIV. Literally Translated by Fathers of the English Dominican Province, Second and Revised Edition* (London, 1922), 242.
[28] For example, "astronomy is not only pleasant, but also very useful to be known: it cannot be denied that this art unfolds the admirable wisdom of God"; John Calvin, "Chapter 1" in *Commentaries on the First Book of Moses, called Genesis, by John Calvin, Vol. 1, John King, trans.* (Edinburgh, 1847) par. 16, 86-87.
[29] *Ibid.*, 86-87.
[30] John Calvin, "Psalm CXXXVI" in *Commentary on the Book of Psalms, by John Calvin, Vol. 5, James Anderson, trans.* (Edinburgh, 1849) par. 7, 184-185.

It seems that, even setting aside Copernicanism, "Corporum Coelestium Magnitudines", with its little moon contra Genesis, would have been controversial and thus eye-catching for some readers, an illustration of something perhaps contrary to their faith, in a way the other illustrations in *Harmonia Macrocosmica* would not. Those illustrations merely show the universe as envisioned under different "hypotheses", or show constellations, or the continents of the Earth, or geometric models for calculating celestial motions (Figure 1). "Corporum" stands out; it seems to offer caution regarding the entire matter of science and assessing things contrary to faith. This question, it reminds the educated reader, has been addressed by Augustine, by Aquinas, and by Calvin. It is difficult to imagine that Ricci, Kircher, and other knowledgeable readers would overlook this diagram.[32]

**Conclusion**

The diagram showing the "Corporum Coelestium Magnitudines" or "Sizes of Celestial Bodies" in Andreas Cellarius's *Harmonia Macrocosmica, seu Atlas Universalis et Novus*, or *Celestial Atlas* provides support to Daniel Stolzenberg's claim that the atlas's publisher and the Roman Inquisition quietly collaborated in 1660 to enable this explicitly Copernican book to circulate in Italy. "Corporum", by masking what many viewed as one of the greatest scientific weaknesses of the Copernican system, was as "blatantly pro-Copernican" as the text of the atlas. Unlike that text, "Corporum" was provided by the publisher to the Inquisition for review prior to publication. The diagram inherently spoke to matters of science and faith that hearkened all the way back to

---

[31] Pascal to Rev. Fr. Annat, S.J. (March 24, 1657), *The Provincial Letters of Pascal* (London, 1847), 369-399, here 393. Pascal also referenced Galileo and the world system debate (395). Another example that shows that the Genesis star size question had not been forgotten is Andrew Willet, *Hexapla in Genesin, that is, a Sixfold Commentary upon Genesis, etc.* (London, 1605) who discusses the smallness of the moon versus the stars on page 10.

[32] The argument of this paper is centered on the images from Cellarius's atlas that Weyerstraet provided to the Inquisition for review, and not on the text, but it is worth noting that the question of the "two great lights" is directly mentioned in the atlas's prologue (p. 9, "Solis, & Lunae, duorum Luminarium Magnorum Sacer Textus mentionem facit [etc.]"). The text accompanying "Corporum" does not mention this subject directly. However, it does provide the estimates of celestial body sizes made by other astronomers than Clavius, showing the lack of agreement among astronomers but emphasizing that they all agree that the sun is large and the moon is not, and that astronomers in this regard fear no censure by the theologians (pars prior p. 71, "Ex his omnibus manifesto patet ingens Astronomorum in definiendis Corporum Mundanorum quantitatibus…. qui Theologicam Censuram non expavescentes [etc.]").

Ptolemy and St. Augustine. It is difficult to imagine the diagram, and the pro-Copernican nature of *Harmonia Macrocosmica*, escaping notice of the Inquisition's reviewers both before the atlas was published and afterwards when the atlas was shipped to Italy for sales. Stolzenberg's claims that "the papacy tacitly permitted the circulation of an explicitly Copernican book at a surprisingly early date" and that "the Copernican *Celestial Atlas* circulated in Italy because of, not despite, the Holy Office" are bold, but a look at the details of this one important illustration from that atlas supports those claims.